\documentclass[11pt,a4paper]{article}

\usepackage{epsfig}
\usepackage{amsmath}
\usepackage{graphicx}
\usepackage{float}
\usepackage{subfig}
\usepackage{vmargin}
\usepackage{mathrsfs}
\usepackage{mathbbol}
\usepackage[parfill]{parskip}
\usepackage[noblocks]{authblk}
\usepackage[citecolor=blue,colorlinks=true,linkcolor=blue]{hyperref}
\usepackage[super,comma]{natbib}

\newcommand{\be}{\begin{equation}}
\newcommand{\ee}{\end{equation}}
\newcommand{\ba}{\begin{equation} \begin{aligned}}
\newcommand{\ea}{\end{aligned} \end{equation}}

\setmargnohfrb{2cm}{2cm}{2cm}{2cm}

\title{\sc Total Effect Analysis of Vaccination on Household Transmission in the 
Office for National Statistics COVID-19 Infection Survey}

\author[1,2,3,*]{Thomas House}
\author[1,3]{Lorenzo Pellis}
\author[4,5]{Emma Pritchard}
\author[6]{Angela R.~McLean}
\author[4,5,7,8]{A.~Sarah Walker}
\affil[1]{Department of Mathematics, University of Manchester, Manchester, UK}
\affil[2]{IBM Research, Hartree Centre, Daresbury, UK}
\affil[3]{The Alan Turing Institute for Data Science and Artificial Intelligence, London, UK}
\affil[4]{Nuffield Department of Medicine, University of Oxford, Oxford, UK}
\affil[5]{The National Institute for Health Research Oxford Biomedical Research
Centre, University of Oxford, Oxford, UK}
\affil[6]{Department of Zoology, University of Oxford, Oxford OX1 3SZ, UK}
\affil[7]{The National Institute for Health Research Health Protection Research
Unit in Healthcare Associated Infections and Antimicrobial Resistance at the
University of Oxford, Oxford, UK}
\affil[8]{MRC Clinical Trials Unit at UCL, UCL, London, UK}
\affil[*]{Corresponding Author: thomas.house@manchester.ac.uk}

\date{}

\begin{document}

\maketitle
\thispagestyle{empty}

\begin{abstract}
\noindent{}We investigate the distribution of numbers of secondary cases in
households in the Office for National Statistics COVID-19 Infection Survey
(ONS CIS), stratified by timing of vaccination and infection in the households.
This shows a total effect of a statistically significant approximate halving of
the secondary attack rate in households following vaccination.
\end{abstract}

\section*{Introduction}

The ongoing COVID-19 pandemic has, at the time of writing, led to over 4
million confirmed deaths worldwide\citep{WHO_Sitrep}. This has in turn caused
Governments to implement significant changes to the way in which societies
function, often including compulsory isolation, implementation of test and
trace systems, and closure of sectors of the economy\citep{hale_global_2021}.
Since they became available, vaccines have been deployed in one of the fastest
ever campaigns with over 3 billion doses administered to
date\citep{WHO_Sitrep}.

In addition to questions about vaccine efficacy on recipient disease
outcomes\citep{Polack:2020,Lumley:2021}, there is a question about the impact
of vaccination on onwards transmission. This was considered by
\citet{Harris:2021} using data from the HOSTED dataset, a passive surveillance
system derived from England's Test and Trace (T\&{}T) system. They reported an
overall secondary attack rate (SAR) in households of 10\%, with adjusted odds
ratios of 0.52 and 0.54 for index case vaccination with ChAdOx1 and BNT162b2
respectively. One potential concern about this study is the biases inherent in
T\&{}T data, and so we seek to see if the estimated vaccine efficacy can be
reproduced under a different study design.

Here, we analyse data from the Office for National Statistics (ONS) COVID-19
Infection Survey (CIS), a large community-based longitudinal household survey
of individuals aged 2 years and older living in randomly selected private
households across the UK\citep{Pouwels:2021}. Due to differences in study
design we take a different analytical approach from that of
\citet{Harris:2021}, but address the same question of the impact of vaccination
on transmission.

\section*{Methods}

\subsubsection*{Data}

In the ONS CIS, households are recruited from the general population and
visited regularly for testing, which is independent of symptoms or vaccine
status.  For visits to be included in the current dataset, participants had to
be aged 16 years or over and have either a positive or negative swab result
from 1st December 2020 to 31 May 2021. We did not differentiate between
vaccines since our aim is to obtain a total effect of the programme as
implemented.

As PCR-positive results may be obtained at multiple visits after infection,
positive tests were grouped into episodes. We defined the start of a new
infection episode as the date of either: (1) the first PCR-positive test in the
study or T\&{}T positive (not preceded by any study PCR-positive test by
definition); (2) a PCR-positive test after four or more consecutive negative
CIS tests; or (3) a PCR-positive test at least 90 days after the start of a
previous infection episode, with one or more negative tests immediately
preceding this.  

Visits were dropped if they were within a positive episode, unless the first
positive in the episode was from T\&T, in which case the first CIS positive (if
any) within that episode was kept in the dataset (as T\&T positives were not
considered as positive outcomes in the dataset).   

Households are stratified into the following three categories:
\begin{itemize}
\setlength\itemsep{-.5em}
\item Positive First \& No Vaccine: First vaccine dose in household received
more than 21 days after first positive episode in household and never
vaccinated households.
\item Intermediate: Difference in time between first vaccine dose in household
and first positive episode in household less than 21 days.
\item Vaccine first: First vaccine dose in household received more than 21
days before first positive episode in household.
\end{itemize}
As we will see, the `intermediate' group is important to ensure that the net
impact of a \emph{completed} vaccination is captured appropriately.

This choice -- i.e.\ stratification by overall household vaccination status --
is necessary because the study design involves testing during a systematically
scheduled visit, meaning the dates of first known positives in households are
often simultaneous and an index case cannot be straightforwardly identified.

\subsubsection*{Analysis}

Here we seek to calculate a \emph{total effect} of having at least one
completed vaccination in a household before introduction of infection, with no
attempt to determine causation, mediation, confounding etc.  We quantify
uncertainty in the results using bootstrapping.

Standard bootstrapping involves repeatedly re-sampling the full dataset with
replacement to quantify uncertainty. Here we are interested in the proportion
of secondary cases generated (the Secondary Attack Rate, or SAR) and the more
detailed distribution of secondary cases in households. If we have $m$
households and the $i$-th household has size $n_i$ and $y_i$ positives, then
let the set of households with at least one infection be $\mathcal{I} =
\{i\in[m]|y_i\geq 1\}$, then the SAR is
\be
\mathrm{SAR} = \frac{\sum_{i\in\mathcal{I}}(y_i - 1)}{\sum_{i\in\mathcal{I}}(n_i-1)} \text{ .}
\ee
We will also be interested in the overall distribution of the $y_i$'s, split
into the three vaccine status groups. To assess uncertainty in these, standard
bootstrapping is not appropriate due to 0\% and 100\% counts, so we calculate
generalised Jeffreys intervals by sampling from the conjugate Dirichlet
distribution to the observed data and then sampling from a multinomial with the
probability vector sampled from the Dirichlet. In each case we use 20,000
bootstraps.

\section*{Results and Discussion}

The SAR estimates and 95\% CIs are as below.
\begin{itemize}
\item Positive First \& No Vaccine: SAR = 23.5[22.6,24.4]\%.
\item Intermediate: SAR = 29.7[22.8,37.1]\%; one-sided p-value for hypothesis
that this is larger than Positive First \& No Vaccine = 0.040.
\item Vaccine first: SAR = 12.5[4.0,23.3]\%; one-sided p-value for hypothesis
that this is larger than Positive First \& No Vaccine = 0.023.
\end{itemize}

The interpretation of these results is that prior vaccination is significantly
associated with lower secondary attack rates in households.  The higher risks in
intermediate households may be related to behaviour, although this would require
further analysis, potentially using the regression methods of \citet{House:2021}.

We now compare with the results from \citet{Harris:2021}; while our overall SAR
is over twice theirs due to different study design, we can determine if the relative
effect is consistent in the following manner. If $\mathrm{OR}$ stands for the odds
ratio in \citeauthor{Harris:2021}, and $\mathrm{SAR}$ for the secondary attack rate in
our positive first and no vaccine group, then the secondary attack rate that would
follow from combination of these two numbers is, after some manipulation of the
definitions of an odds ratio in logistic regression and the secondary attack rate,
\be
Q = \frac{\mathrm{SAR}\times \mathrm{OR}}{\mathrm{SAR}\times \mathrm{OR} + (1-\mathrm{SAR})}
\text{ .}
\ee
For the ChAdOx1 estimate in \citeauthor{Harris:2021} we obtain $Q \approx
13.8[8.5,19.3]\%$, and for the BNT162b2 estimate, $Q \approx
14.2[10.3,18.4]\%$. Both are consistent with our vaccine first group SAR
estimate, meaning that both study designs are consistent in terms of the
inferred relative secondary attack rate following vaccination.

\section*{Acknowledgements}

The ONS CIS is funded by the Department of Health and Social Care with in-kind
support from the Welsh Government, the Department of Health on behalf of the
Northern Ireland Government and the Scottish Government.  TH is supported by
the Royal Society (grant number INF/R2/180067). LP is supported by the Wellcome
Trust and the Royal Society (grant number 202562/Z/16/Z). TH and LP are also
supported by the UK Research and Innovation COVID-19 rolling scheme (grant
numbers EP/V027468/1, MR/V028618/1 and MR/V038613/1) as well as the Alan Turing
Institute for Data Science and Artificial Intelligence. EP and ASW are
supported by the National Institute for Health Research Health Protection
Research Unit (NIHR HPRU) in Healthcare Associated Infections and Antimicrobial
Resistance at the University of Oxford in partnership with Public Health
England (PHE) (NIHR200916). EP is also supported by the Huo Family Foundation.
ASW is also supported by the NIHR Oxford Biomedical Research Centre, by core
support from the Medical Research Council UK to the MRC Clinical Trials Unit
(MC\_UU\_12023/22), and is an NIHR Senior Investigator. The authors would like
to thank the ONS CIS team as well as Arturas Eidukas and Kaveh Jahanshahi from
the ONS Data Science Campus project support.

\clearpage

\section*{Figures}

\begin{figure}[H]
\centering
\includegraphics[width=0.5\textwidth]{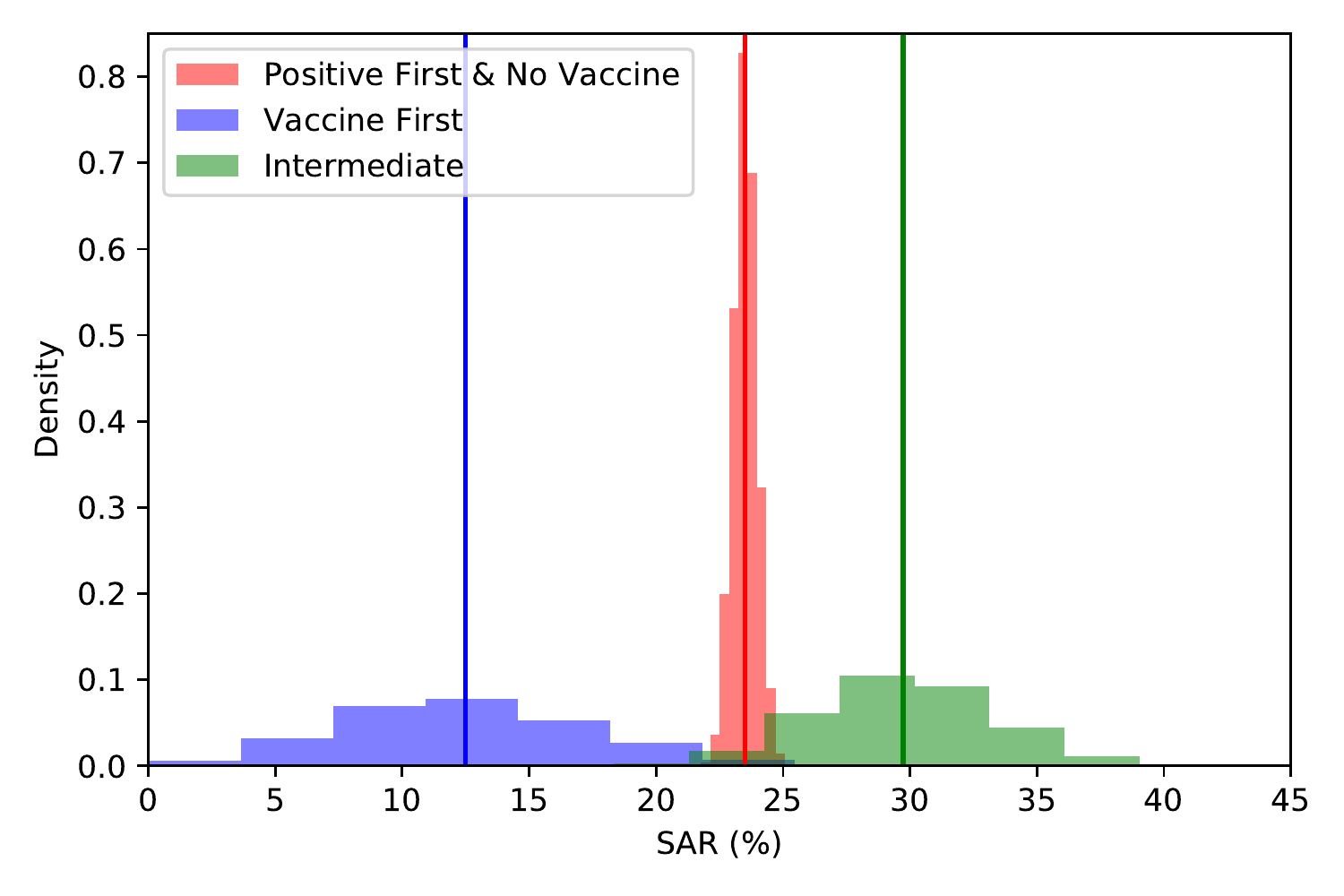}
\caption{Household secondary attack rates (SARs) bootstrapped at the
whole-dataset level.}
\label{fig:sar}
\end{figure}

\begin{figure}[H]
\centering
\includegraphics[width=\textwidth]{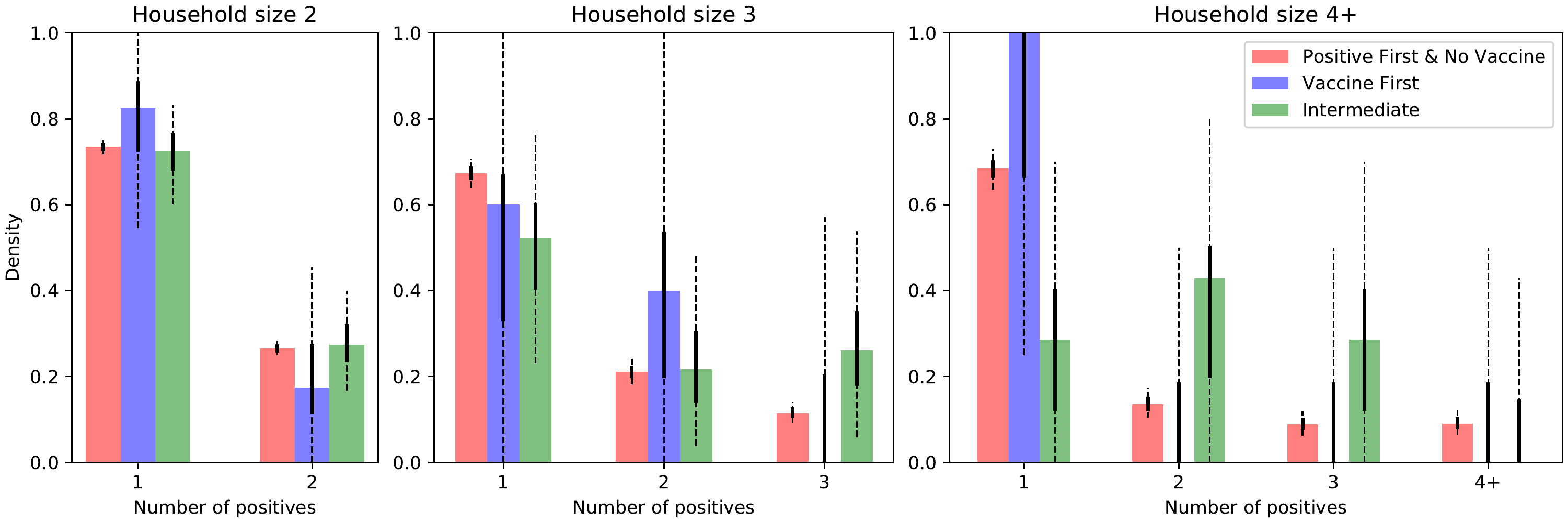}
\caption{Histograms of numbers positive in households stratified by
household sizes with 50\% and 95\% CIs from whole-sample parametric
bootstrapping shown.}
\label{fig:hist}
\end{figure}

\end{document}